\documentclass{article}  

\usepackage{graphicx}

\usepackage{geometry}
\usepackage{amsmath}
\usepackage{graphicx}
\usepackage{natbib}
\usepackage{color}
\usepackage{lineno}
\usepackage{amstext}
\usepackage{multirow}
\usepackage{ulem}
\usepackage{lscape}
\usepackage{txfonts}
\usepackage{subfigure}
\usepackage{float}
\usepackage{ctable} 
\usepackage[bottom]{footmisc}
\usepackage{caption}  
\usepackage{makecell}
\usepackage[colorlinks=true, linkcolor=blue, citecolor=blue, urlcolor=blue]{hyperref}

\usepackage{etoolbox}
\makeatletter
\patchcmd{\@afterheading}{\@nobreaktrue}{}{}{}
\makeatother

\newcommand{\jgr}{J. Geophys. Res. }
\newcommand{\grl}{Geophys. Res. Lett. }
\newcommand{\icarus}{Icarus }
\newcommand{\aap}{Astron. Astrophys. }
\newcommand{\apj}{Astrophys. J. }
\newcommand{\apjl}{Astrophys. J. Lett. }

\newcommand{\aj}{Astron. J. }
\newcommand{\planss}{Planet. Space Sci. }
\newcommand{\araa}{Ann. Rev. Astron. Astrophys. }
\newcommand{\ssr}{Space Sci. Rev. }
\newcommand{\mnras}{Mon. Not. R. Astron. Soc. }
\newcommand{\nat}{Nature }

\newcommand{\jchemphys}{J. Chem. Phys. }

\newcommand{\psj}{Plan. Sci. J. }

\newcommand{\arcsec}{\hbox{$^{\prime\prime}$}}

\newcommand{\degre}{\ensuremath{^\circ}}

\begin{document} 


\begin{center}
\textbf{\LARGE New $^{12}$C/$^{13}$C and $^{14}$N/$^{15}$N isotopic ratio measurements in Jupiter's stratosphere revealed by ALMA}\\
\end{center}
\vspace{0.4cm}
\normalsize

\large\noindent C. Lefour$^{1}$, T. Cavali\'e$^{1,2}$, R. Moreno$^{2}$, L. Rezac$^{3}$, T. Fouchet$^{2}$, E. Lellouch$^{2}$, P. Hartogh$^{6}$\normalsize\\
\vspace{0.2cm}

\noindent$^1$Laboratoire d'Astrophysique de Bordeaux, Univ. Bordeaux, CNRS, B18N, all\'ee Geoffroy Saint-Hilaire, 33615 Pessac, France (ORCID: 0009-0004-6922-9388, email: camille.lefour@u-bordeaux.fr)\\
$^2$LIRA, Observatoire de Paris, Université PSL, CNRS, Sorbonne Universit\'e, Universit\'e Paris Cité, 5 place Jules Janssen, 92195 Meudon, France\\
$^3$Max-Planck-Institute for Solar System Research, G\"ottingen, Germany\\

\vspace{0.2cm}
\noindent\textbf{Received:} 2 February 2026\\
\noindent\textbf{Accepted:} 16 May 2026\\
\noindent\textbf{Published:} ?? ?? 2026\\
\vspace{0.2cm}

\noindent\textbf{DOI: https://doi.org/10.1051/0004-6361/202659264} \\
\vspace{0.5cm}

\section*{Abstract}
The collision of comet Shoemaker-Levy 9 (SL9) with Jupiter in 1994 changed the chemical composition of the Jovian stratosphere for decades. New molecules were detected minutes after the impacts (HCN, CO, H$_2$O, CS, etc.) and some are still present today. They were deposited in the stratosphere at pressures lower than 0.1 mbar and were most probably formed by shock-induced chemistry recombining Jovian and cometary material. However, the question of the origin of these molecules is still not completely understood. One way to address this open question is to determine the isotopic composition of the new molecules. Isotopic ratios have long been measured in the Solar System. They present a variety of values depending on the object or the molecule and therefore trace different reservoirs of material. Derivations of carbon and nitrogen isotopic ratios in HCN four years after SL9 showed atypical depletions in the heavier isotopes that had never been observed before in the Solar System. These results suggested an unusual cometary composition or an unknown fractionation mechanism in the hot and shocked air parcels.
We aim to measure carbon and nitrogen isotopic ratios in HCN to shed light on the puzzling results of 1998.
With Atacama Large Millimeter/submillimeter Array data from 2017 and radiative-transfer calculations, we derived the abundance of two HCN isotopologues, H$^{13}$CN and HC$^{15}$N, at pressures probed from 0.03 to 1.8 mbar.
We find $^{12}$C/$^{13}$C $= 73\pm5$ and $^{14}$N/$^{15}$N $ = 245\pm29$, respectively (0.76-0.87) and (0.80-1.00) times the terrestrial references, and (0.69-0.87) and (0.42-0.70) times the solar--Jovian bulk values. In contrast to the strong depletions reported in 1998, our values are instead compatible with an enrichment in the heavier isotopes relative to the Jovian bulk.
We interpret these enrichments as the direct signature of the cometary contribution in HCN and/or as 23 years of chemical evolution in the Jovian atmosphere.

\section{Introduction}

Jupiter and the other giant planets are constantly subject to external modifications that enrich their atmosphere with new material (e.g. \citealt{Lellouch1995,Hesman2007,Hartogh2011a,Cavalie2010,Cavalie2014,Moses2017}). A major example consisted of the 21 fragments of the comet Shoemaker-Levy 9 (SL9), which impacted the Jupiter atmosphere in July 1994 at 44\degre S during a week \citep{Noll1996book}. Molecules that were previously undetected in the stratosphere of Jupiter (H$_2$O, CS, HCN, OCS, and many more) were observed minutes, months (see the review of \citealt{Lellouch1996book}), and, for the most long-lived, years after the impacts \citep{Lellouch2002,Lellouch2006,Moreno2003,Iino2016,Cavalie2013,Cavalie2023b,Benmahi2020,Rodriguez-Ovalle2025}. Other molecules that were already present before SL9, such as CO \citep{Bezard2002}, saw their stratospheric abundance enhanced after the impacts \citep{Lellouch1997}. These molecules are the direct tracers of the dynamical and chemical processes triggered by the impacts \citep{Zahnle1996,Lellouch1996book}. Therefore, their study during and after the impacts can make it possible to reconstruct their formation and/or deposition history and their further temporal and spatial evolution in the Jovian atmosphere.

The question of the origin of these molecules, i.e. from which reservoir of material (Jovian or cometary) and in which conditions (explosion, fireball, splashback, etc.) they were formed, is still open as it involves many parameters. Firstly, the elementary composition of the comet was not known before the impacts \citep{Crovisier1996} and some hints can only be found during or after the impacts in the new material formed/deposited in the Jovian atmosphere \citep{Lellouch1996book}. However, the cometary material completely mixed with the Jovian air in the impact fireball, making it difficult to separate the cometary from the Jovian contribution in the newly formed species \citep{Lellouch1995,Crovisier1996}. In addition, the different phases of impact (fragment entry, final explosion, rising fireball, plumes fallback) led to different physical processes \citep{Zahnle1996} and thus altered the atmosphere differently depending on the altitude and horizontal extent at which they occurred. For example, CS, CO, and HCN were probably formed by shock-induced chemistry \citep{Zahnle1996}, recombining cometary and Jovian material in the extreme conditions of the fireball and plumes splashback in which temperatures reached several thousands of kelvins \citep{Chapman1996book,Bjoraker1996,Nicholson1996book}. They were all deposited during the plumes' re-entry into the stratosphere at pressures of 0.05--0.2 mbar \citep{Lellouch1997}. In this case, it becomes difficult to decipher the relative contributions of cometary and Jovian materials to the final products \citep{Lellouch1996book}. 
On the other hand, the detection of NH$_3$ and H$_2$S (normally only present in Jupiter's troposphere) in the lower stratosphere suggested that convection driven by the heat released at a few bars by the fragment explosion overshoot to the lower stratosphere and transported here Jovian air and cloud materials \citep{Yelle1996,Griffith1997}. In this case, the lower stratospheric NH$_3$ and H$_2$S probably have a purely Jovian origin.

In addition, after the formation and deposition of the species at a given altitude, their initial distribution was modified as a function of time by dynamics and chemistry. The photolysis of the short-lived species (e.g. NH$_3$) led with time to the enrichment in other species (e.g. HCN; \citealt{Moreno2003,Lellouch2006,Moses1996book}), thus altering the primordial composition of the new species and making it even more difficult to assess the origin of their material. Also, at first confined to the impact sites, the species spread out with longitude within a few months after the impacts and with latitude to cover the entire disc years later \citep{Moreno2003,Lellouch2006,Cavalie2023b}.

In this context, measuring the isotopic composition of the new stratospheric species can still provide insights into their formation processes and their dominant sources of elements (Jovian or cometary), but the isotopic ratios can also diagnose the temporal evolution of the exogenic molecules from interactions in the Jovian atmosphere. Isotopic ratios are often used to trace the origin of a material because they reflect the physical and chemical conditions in which their bearing molecule was formed \citep{Nomura2023}. They are usually compared to solar values \citep{Marty2011,Lyons2018} or to the Earth references \citep{IAEA1995,Coplen1992} to reveal any fractionation processes as a function of the type of object, the distance to the Sun, or the molecule \citep{Nomura2023}. Only the carbon, nitrogen, and sulphur isotopic ratios were estimated four years after SL9 impacts in exogenic HCN and CS by \citet{Matthews2002}. They derived an abnormal depletion in all three heavier isotopes ($^{13}$C, $^{15}$N, $^{34}$S), with $^{12}$C/$^{13}$C, $^{14}$N/$^{15}$N, and $^{32}$S/$^{34}$S ratios nominally higher by factors of three, ten, and three, respectively, than the Earth references of 89, 272, and 23.

Isotopic ratios have been estimated in a wide range of Solar System objects and in several of their molecules over the past 50 years. For example, there is a general trend that the $^{12}$C/$^{13}$C ratio is quasi-constant throughout the Solar System at a value close to solar and terrestrial, independently of the object (meteorites, comets, giant, and terrestrial planets) or the molecule \citep{Nomura2023,Woods2009,Woods2009DB}. This absence of notable isotopic fractionation seems to indicate that the Solar System objects accreted carbon from a common reservoir, or that all carbon reservoirs present the same isotopic composition \citep{Fletcher2009a,Nomura2023}. A departure from this common trend, as reported by \citet{Matthews2002} in the carbon present in the SL9-produced HCN, is thus very intriguing. On the other hand, the nitrogen isotopic composition presents a broader range of values in the Solar System and is highly dependent on the type of the object. In the giant planet atmospheres, the $^{14}$N/$^{15}$N ratio is close to solar but is super-terrestrial by a factor of 1.6 \citep{Owen2001,Fouchet2000,Fletcher2014}. Comets are the most $^{15}$N-enriched objects of the Solar System. Their $^{14}$N/$^{15}$N nominal values are at least a factor of two lower than the terrestrial reference and a factor of three lower than the solar and Jovian values \citep{Hily-Blant2017}. The dispersion of values in the nitrogen isotopic ratio among Solar System objects is thus helpful to decipher the origin of the objects, as opposed to the carbon ratio. However, as for carbon, the strong depletion in $^{15}$N in HCN after SL9 completely departs from any Solar System measurements, even from the Jovian bulk, which is the least enriched. Unfortunately, no complementary observations of the isotopic ratios in Jupiter's exogenic molecules have been performed these past 30 years. The results of \citet{Matthews2002} thus remain puzzling and no definitive answer has been found regarding the origin of the elements composing HCN.

In this work, we used 2017 Atacama Large Millimeter/submillimeter Array (ALMA) observations of HCN and two of its isotopologues, H$^{13}$CN and HC$^{15}$N, to estimate the carbon and nitrogen isotopic ratios in HCN in the Jupiter stratosphere 23 years after SL9 impacts, and to try to understand the intriguing results of \citet{Matthews2002} in 1998. We detail the dataset in Section \ref{part:Observations} and the radiative-transfer model used to simulate the observations in Section \ref{part:Models}. We then present the results in Section \ref{part:Results} and discuss and compare them to other measurements in Section \ref{part:Discussion}. We give our conclusions in Section \ref{part:Conclusion}.

\section{Observations}
\label{part:Observations}

We used the ALMA Cycle 4 dataset already presented in \citet{Cavalie2021,Cavalie2023b} from which they derived, respectively, stratospheric wind measurements and vertical abundance profiles from the CO (J=3--2) line at 345.7960 GHz and the HCN (J=4--3) line at 354.5055 GHz. General information on the dataset is presented here, but more details on the observation setup and data reduction can be found in \citet{Cavalie2021,Cavalie2023b}.

The Jupiter observations were recorded on March 22, 2017 as part of the 2016.1.01235.S project. The ALMA configuration at the time of the observations consisted of 42 antennas with a diameter of 12 m and baselines extending from 15.1 to 160.7 m. The resulting beam size is $1.0-1.2$\arcsec\, (east-west and north-south) with a position angle of 87.3\degre. The complete dataset consists of four different spectral cubes of Jupiter, taken in four spectral bands, from which we extracted two smaller cubes at 345.3398 and 344.2003 GHz encompassing, respectively, the H$^{13}$CN and HC$^{15}$N (J=4--3) lines. These are the only isotopic lines present in the dataset. The continuum level was subtracted from these cubes. The spectral resolution is $\Delta \nu = 0.488$ MHz, resulting in a spectral resolving power of about $7 \times 10^5$ at 344-345 GHz. 

Jupiter's angular diameter was 43.8\arcsec\, at the equator, and its rotation axis was tilted by 23.7\degre\footnote{Jupiter's geometry was found in the JPL/Horizons database at \url{https://ssd.jpl.nasa.gov/horizons/}.}. The sub-observer latitude was $-$3.5\degre\, and the sub-observer longitude was 65.4\degre W\footnote{The latitudes are given in the planetocentric geometry and the longitudes in System III.} at the beginning of the observation. During the 24 min on-source integration time, the planet rotated by 15\degre; we thus observed the planetary western limb in a longitude range spanning 155--170\degre W and 335--350\degre W at the eastern limb.

The small beam size compared to the planet's disk resulted in a latitudinal resolution of 2-3\degre\, per beam at the low latitudes and up to 8\degre\, per beam at the higher latitudes. From the knowledge of Jupiter's geometry and using the continuum map of Jupiter, we found a pointing offset of $\Delta$RA=0.053\arcsec\, and $\Delta$DEC=$-$0.039\arcsec\, and re-centred the planetary disc accordingly. For both spectral cubes (H$^{13}$CN and HC$^{15}$N), we oversampled the beam by a factor of four and extracted 540 spectra at the planetary limb, which is defined by the projected planetary ellipse at the 1 bar level. We only extracted spectra at the limb where we have the highest line sensitivity. 

\begin{figure*}[!ht]
\centering
\includegraphics[width=\textwidth]{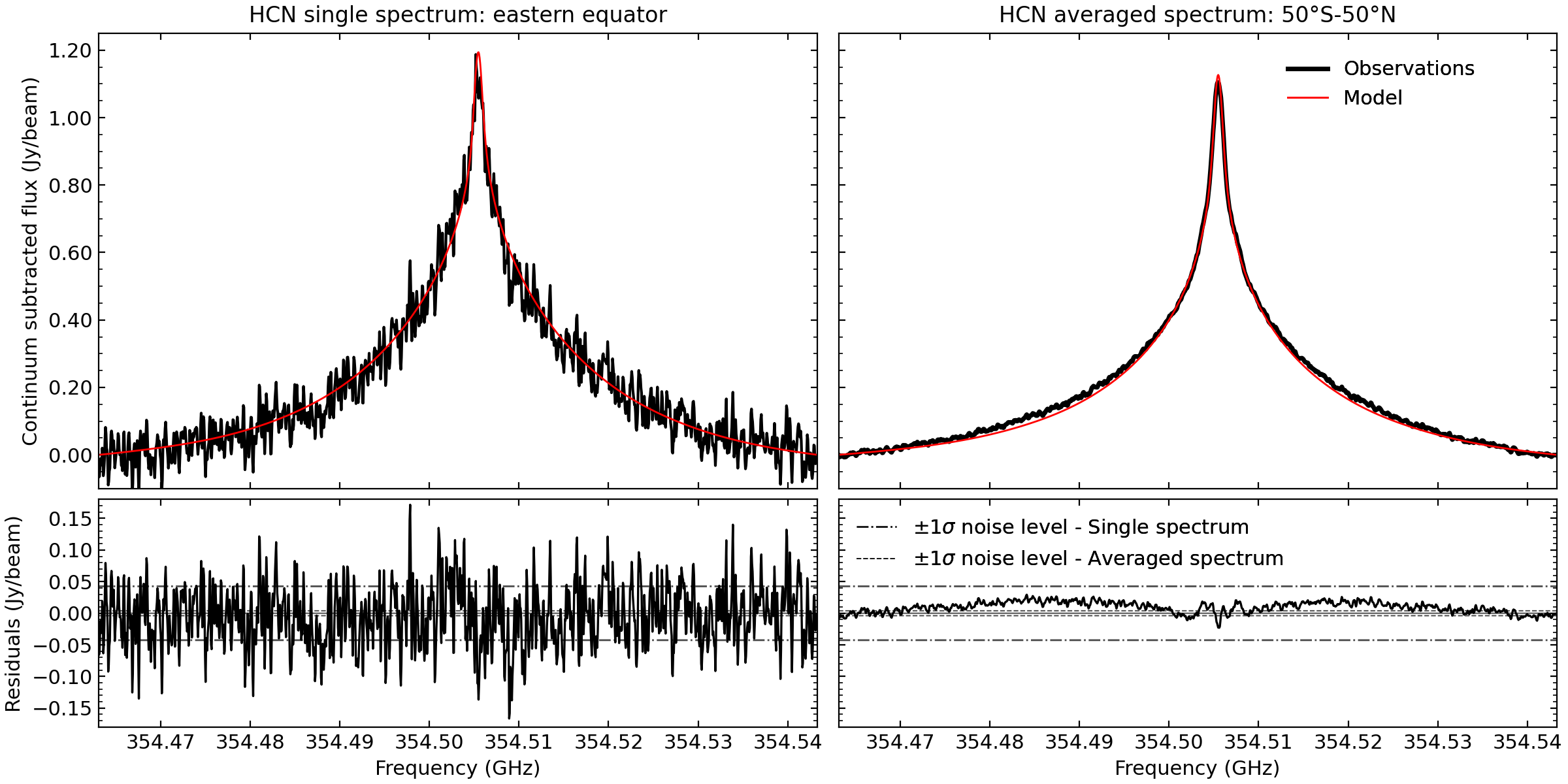}
\caption{HCN limb spectra. (\textit{left}) HCN limb spectrum extracted at a single position on the eastern equatorial limb at 3\degre N. (\textit{right}) HCN spectrum averaged over the two limbs and on the latitudinal range 50\degre S-50\degre N. The observed spectra are presented in black and the modelled spectra in red. The differences between the observed and the modelled spectra are shown below with the residuals. The 0 Jy/beam level is depicted by the thin solid black line. The dashed-dotted lines represent the $1\sigma$ noise level in the single spectrum on the left panel, and the dashed lines the $1\sigma$ noise level from the right panel's averaged spectrum. For the single pointing spectra, the continuum level was subtracted and the lines were centred at the rest frequency. The averaged spectra follow the same averaging method described in Section \ref{part:Observations}. Observations come from the dataset presented in \citet{Cavalie2023b}. Modelled spectra were computed with the procedure detailed in Section \ref{part:Models}.}
\label{fig:fit_HCN}
\end{figure*}

\begin{figure*}[!ht]
\centering
\includegraphics[width=0.8\textwidth]{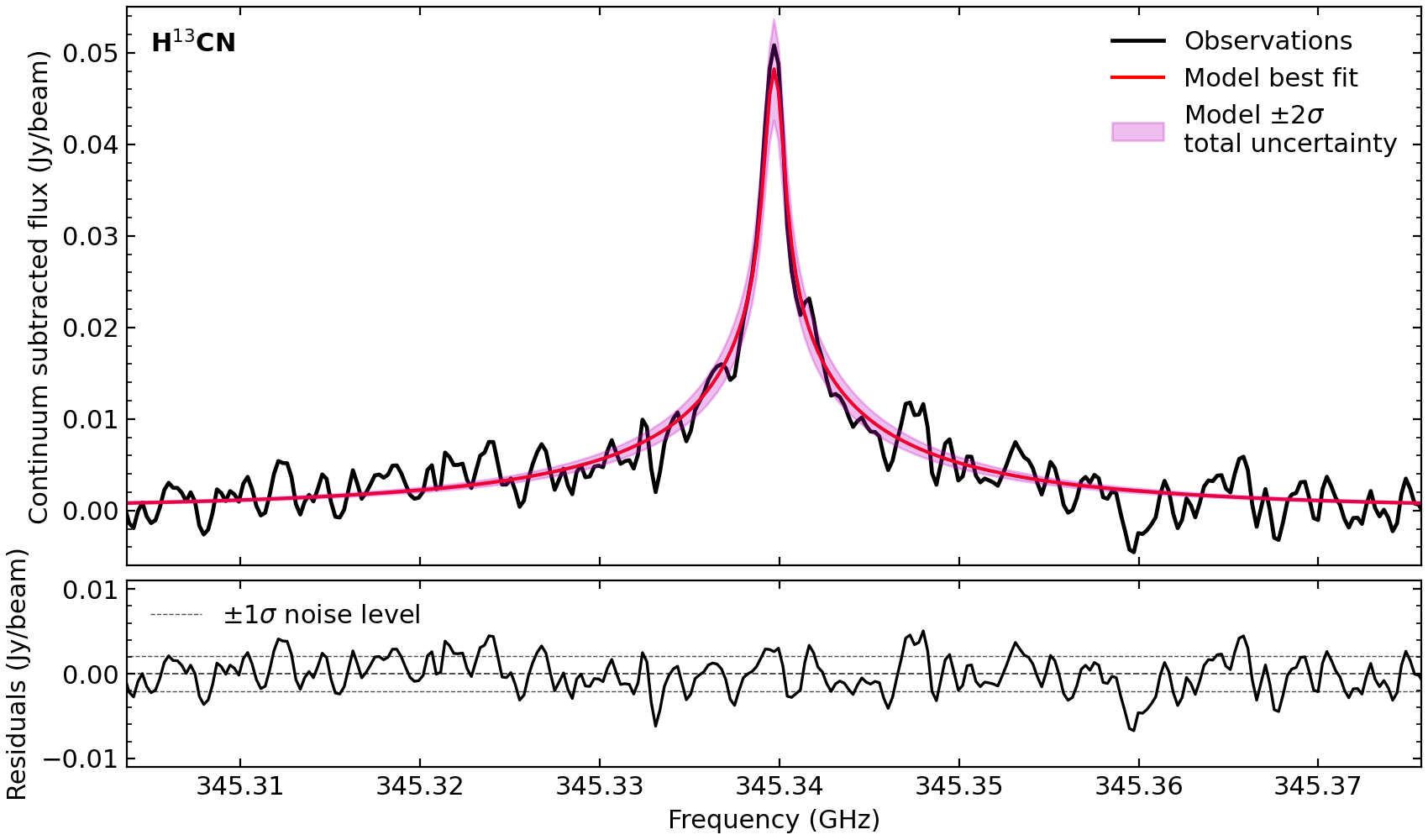}
\caption{(\textit{top}) Observed and modelled spectra of the H$^{13}$CN line at 345.3398 GHz. The observed spectrum is depicted by the black line and corresponds to a latitudinal average on the 50\degre S-50\degre N range from both limbs. The modelled spectrum that best fits the observations is represented by the solid red line. The pink area shows the $\pm2\sigma$ uncertainties on the modelled line. (\textit{bottom}) Residuals corresponding to the difference between the observed spectrum and the model's best fit. The horizontal dotted black lines represent the $\pm 1\sigma$ noise level in the averaged spectrum.}
\label{fig:Resultats_H13CN}
\end{figure*}

\begin{figure*}[!ht]
\centering
\includegraphics[width=0.8\textwidth]{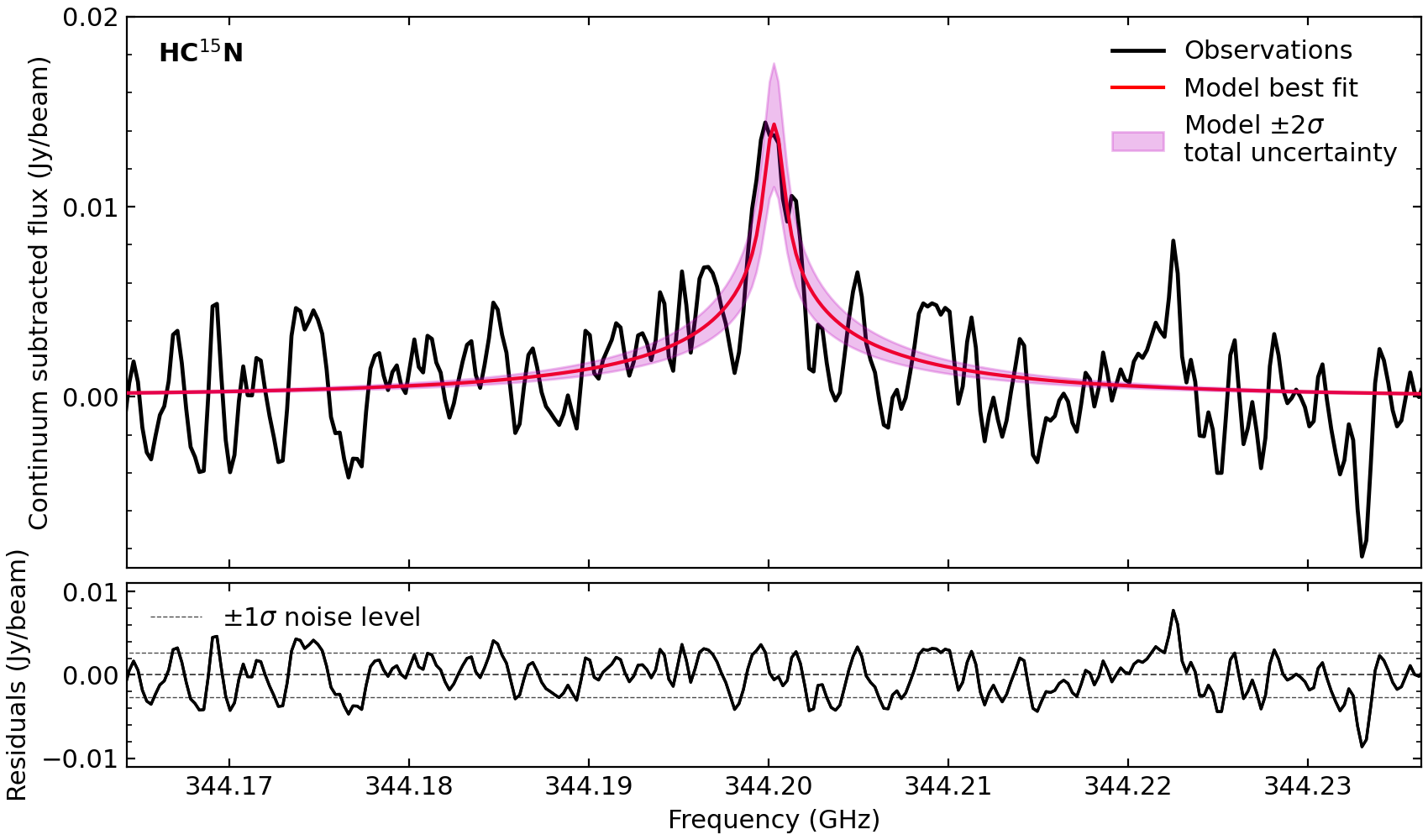}
\caption{Same as Fig. \ref{fig:Resultats_H13CN} but for the HC$^{15}$N line at 344.2003 GHz.}
\label{fig:Resultats_HC15N}
\end{figure*}

The same procedure was performed by \citet{Cavalie2023b} to extract HCN spectra at the limb. Because the spatial resolution is slightly better at the HCN (J=4--3) transition, they extracted a total of 557 limb spectra. The total 24 min observation time was sufficient to detect the HCN line in each of these individual limb spectra and reach signal-to-noise ratios (SNR) of the order of 25 per beam. An example of an HCN individual spectrum is presented in Fig. \ref{fig:fit_HCN}. However, because H$^{13}$CN and HC$^{15}$N are significantly less abundant than H$^{12}$C$^{14}$N, the resulting lines are too faint to be detected in an individual limb-extracted spectrum. Therefore, to increase the SNR, we needed to co-add spectra, and we adopted the following procedure: 1) for each spectrum and on both limbs, we subtracted the Doppler shift caused by the beam-convolved planet rotation to re-centre all the lines at the line rest frequency; then 2) we averaged the spectra on the selected latitudinal range that gives the best SNR; and 3) we subtracted any continuum baseline left with the averaging step. The best latitudinal range was found to be 50\degre S-50\degre N (i.e. 303 selected spectra) and corresponds to the highest and rather constant HCN column density with latitude \citep{Cavalie2023b}. The final spectra obtained with this method are illustrated in Figs. \ref{fig:fit_HCN} (right), \ref{fig:Resultats_H13CN}, and \ref{fig:Resultats_HC15N} for HCN, H$^{13}$CN, and HC$^{15}$N, respectively. We detected the H$^{13}$CN and HC$^{15}$N lines with an SNR at a line peak of 24 and five, respectively, and with noise levels at $1\sigma$ of 2.1 and 2.6 mJy/beam.

\section{Modelling}
\label{part:Models}

We used the radiative-transfer model detailed in \citet{Cavalie2019} (and references therein) to simulate the HCN, H$^{13}$CN, and HC$^{15}$N lines in the same atmospheric, geometric, and instrumental conditions of the observations. The most important code inputs are summarised in Section \ref{part:RT_code}. As the isotopologue lines are optically thin (the opacities at the line centre are 0.015 and 0.004, respectively), the line amplitude is sensitive to the atmospheric temperature and to the abundance of the species. We detail the 3D fields employed in the modelling in Sections \ref{part:Thermal_field} to \ref{part:H13CN_HC15N_field}.

\subsection{Radiative-transfer model}
\label{part:RT_code}

The radiative transfer model from \citet{Cavalie2019} is a line-by-line model accounting for Jupiter's 3D ellipsoidal geometry in the same conditions as the observations (see Section \ref{part:Observations}). The background composition of H$_2$, He, CH$_4$, NH$_3$, and PH$_3$ used for the continuum is the same as in \citet{Cavalie2023b}. 

The pressure broadening coefficient, $\gamma$, and the exponent for the temperature dependency, denoted $n$, were not measured for the H$^{13}$CN and HC$^{15}$N (J=4--3) transitions. \citet{Rohart1987} only gave values for the J=1--0 transition of HC$^{15}$N: $\gamma=0.1308$ cm$^{-1}$.atm$^{-1}$ and $n=0.752$. Here for consistency, in the nominal model, we adopted the pressure-broadening parameters of the H$^{12}$C$^{14}$N (J=4--3) transition from \citet{Rohart1987} (and used by \citealt{Cavalie2023b}) for both heavier isotopologues, i.e. $\gamma=0.1447$ cm$^{-1}$.atm$^{-1}$ and $n=0.756$. In addition, we rescaled the HC$^{15}$N (J=1--0) parameters using the H$^{12}$C$^{14}$N (J=1--0) and (J=4--3) values to estimate the parameters at the HC$^{15}$N (J=4--3) transition: $\gamma=0.1305$ cm$^{-1}$.atm$^{-1}$ and $n=0.770$. The difference implied on the derived HC$^{15}$N abundance by choosing either these rescaled values or the nominal ones is accounted for in the uncertainties. In the absence of information on the H$^{13}$CN parameters, the same approach as for HC$^{15}$N was performed for H$^{13}$CN.

We computed spatially and spectrally convolved spectra of H$^{13}$CN and HC$^{15}$N at each of the 303 selected positions extracted from the limb, contained between 50\degre S and 50\degre N. The modelled continuum level was removed by subtracting a linear slope under the lines. Modelled spectra were then corrected for the beam-convolved planet rotation and co-added following the same method as for the observations, on the same latitudinal range.

\subsection{Thermal field}
\label{part:Thermal_field}

The latitude-pressure thermal field used in this paper is presented in more detail in \citet{Cavalie2023b}. The data were retrieved from IRTF/TEXES observations \citep{Cosentino2017,Sinclair2023}, which were taken near-simultaneously to the ALMA dataset. The thermal vertical profiles inverted from these data span 1 $\mu$bar to 1 bar with a 2 K uncertainty \citep{Cosentino2017}. The final thermal field consists of two 2D latitude-pressure fields (one at each planetary limb) re-gridded on a 1\degre\, latitude grid (see Extended Data Fig. 1 in \citealt{Cavalie2023b}).

\subsection{H$^{12}$C$^{14}$N abundance field}
\label{part:HCN_field}

\begin{figure}[!ht]
\centering
\includegraphics[width=0.49\textwidth]{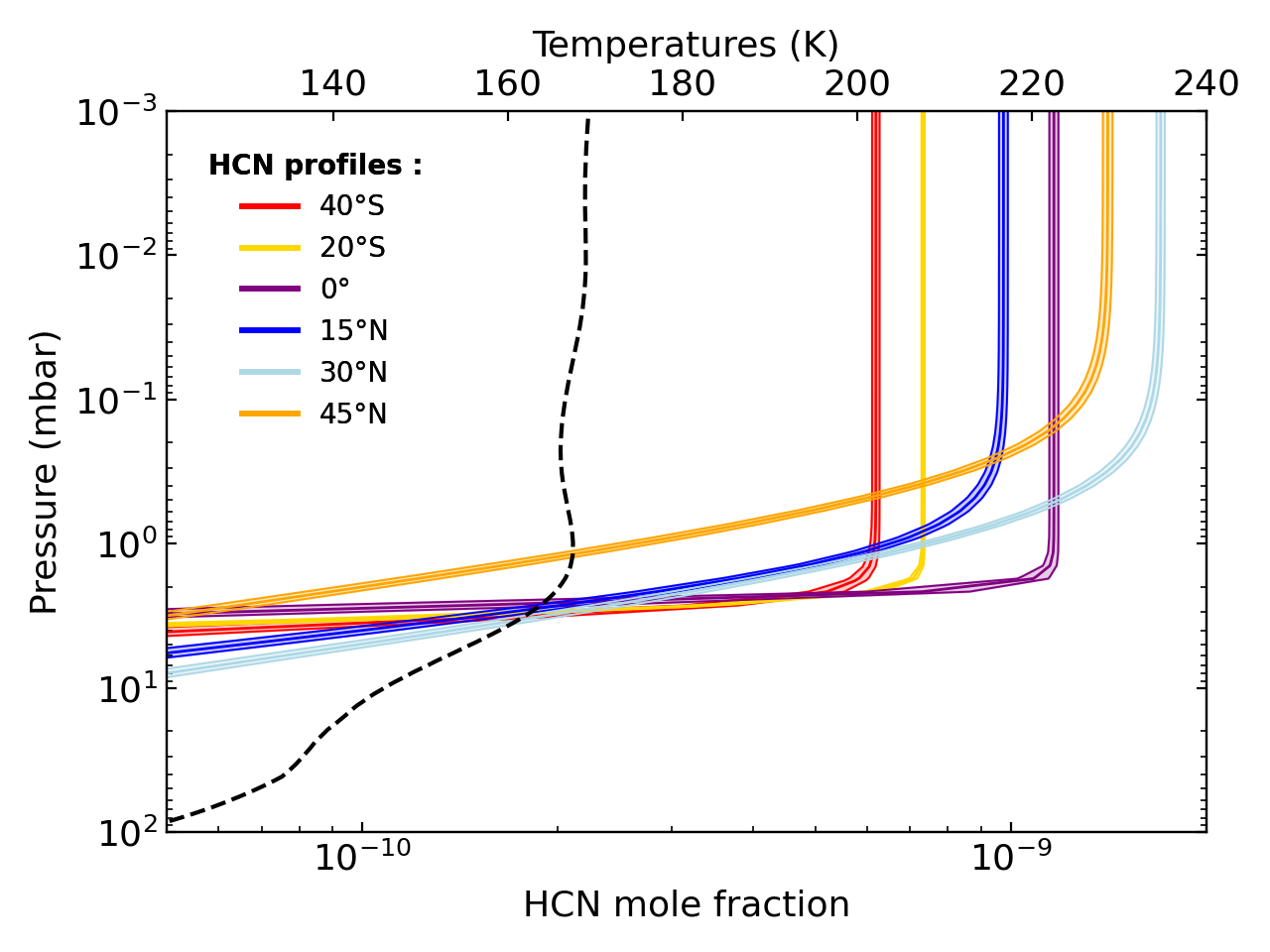}
\caption{Selection of H$^{12}$C$^{14}$N vertical profiles used in our radiative-transfer modelling at different latitudes and all (arbitrarily) taken from the eastern limb. They illustrate the limited variety of profiles retrieved at all low-to-mid latitudes on both limbs, as shown in \citet{Cavalie2023b}. The filled areas show the $1\sigma$ envelope resulting from the $1\sigma$ uncertainties on the retrieved profile parameters. A typical temperature profile extracted at the equator is also displayed on the top x-axis as an example (dashed black line). The full 3D temperature field can be found in \citet{Cavalie2023b}.}
\label{fig:3D_fields_HCN}
\end{figure}

We used the H$^{12}$C$^{14}$N latitude-pressure abundance field determined by \citet{Cavalie2023b} from the same ALMA dataset of March 22, 2017, at both limbs. Their retrievals were performed at 557 positions, which were extracted at the planetary limb from a beam oversampling of a factor of four. The HCN observations benefit from a high spectral resolving power of $3 \times 10^6$ that enables us to resolve the true line shape, which in turn allows us to retrieve the vertical distribution of HCN. Each of their 557 HCN spectra were fitted with an inversion algorithm coupled to the radiative-transfer model presented above, in which the HCN vertical profile was parametrised with three parameters, $a_1$, $a_2$, and $a_3$, and the following relationship:

\begin{equation*}
    y_{\mathrm{HCN}}(z) = a_1 \times \left( 1 + \tanh\left(\frac{z-a_2}{2a_3}\right) \right),
\end{equation*}

\noindent where $y_{\mathrm{HCN}}$ is the HCN abundance or volume mixing ratio (VMR), $z$ is the altitude in km above the 1-bar level, $a_1$ is half the asymptotical high-altitude HCN VMR, $a_2$ is a cut-off level between the upper and lower stratosphere, and $a_3$ indicates the steepness of the slope at the cut-off level, with a stronger slope for larger $a_3$ values. Examples of these retrieved HCN vertical profiles are presented in Fig. \ref{fig:3D_fields_HCN}. The uncertainties on the retrieved vertical profiles come from the combination of several sources: (i) flux-calibration uncertainties, which are reported at a level of 5\%; (ii) retrieval uncertainties on the $a_1$, $a_2$, and $a_3$ parameters resulting from the spectral noise and which translate to an approximately 5\% uncertainty on the retrieved abundances. It should be noted that, even though the HCN line has an opacity of 1.6 at the line centre in the low-to-mid latitudes, any line saturation is accounted for when retrieving the uncertainties on the vertical profile parameters; (iii) uncertainties resulting from the 2 K uncertainty on the temperature field, which translate into a 1\% uncertainty on the abundances; (iv) the uncertainties resulting from the continuum subtraction in the uv-plane during the data analysis. This was reported by \citet{Cavalie2023b} to be the largest source of uncertainties on the retrieved abundances, causing a 30\% uncertainty on the abundances. Considering all four contributions, there is an overall 40\% uncertainty on the HCN vertical profiles.

With the individual HCN vertical profiles of \citet{Cavalie2023b}, the final HCN abundance field used in our radiative-transfer calculations consists on two 2D latitude-pressure fields (one at each limb), re-gridded with a 1\degre\, latitude step, as for the thermal field. This was done by co-adding vertical profiles located within each 1\degre\, latitude step. Because the latitudinal resolution is 2-3\degre\, in one beam in these observations, and because the beam was oversampled by a factor of four in the HCN profile retrieval process, it means that each 1\degre\, latitude step of our grid averages two retrieved profiles on average.

To ensure that the re-gridded HCN field presented above was valid, we ran radiative-transfer calculations for the HCN line at each of the 557 positions. As an example, Fig. \ref{fig:fit_HCN} shows a model-observation comparison for the east-equatorial spectrum. After applying the spectrum-averaging method described in Section \ref{part:Observations}, we obtained the result shown in Fig. \ref{fig:fit_HCN} right. The co-addition of individual spectra results in a significant decrease of the noise level, from 0.043 mJy/beam on a single spectrum, to 0.004 mJy/beam on the averaged one. As a consequence, the mismatches that lied within the noise on individual spectra result in differences after the averaging stage that then become more significant (see the residual on Fig. \ref{fig:fit_HCN}, right). However, those differences remain very limited. This demonstrates that the HCN abundance field can be used as a reference to derive the isotopologue abundances and consequently compute the $^{12}$C/$^{13}$C and $^{14}$N/$^{15}$N isotopic ratios.

\subsection{H$^{13}$CN and HC$^{15}$N abundance field}
\label{part:H13CN_HC15N_field}

We assume H$^{13}$CN and HC$^{15}$N have the same vertical and horizontal distribution as the 3D H$^{12}$C$^{14}$N abundance field detailed above, but rescaled by a constant factor representing the isotopic ratio. Fractionation from condensation with altitude was excluded for two reasons: 1) the HCN vapour pressure curve (see \citealt{Jancso1974,Fray2009,Hudson2023}) indicates that the molecule is far from condensing under Jovian stratospheric temperature and pressure conditions (about 170 K at 0.1 mbar), with mole fractions of the order of $10^{-11}-10^{-9}$; 
and 2) vapour pressure curves between isotopologues are typically very similar. Saturation curves from \citet{Jancso1974} (and references therein) indeed indicate a difference of the order of 10\% between the HCN and HC$^{15}$N curves, which is still very far from the Jovian atmospheric conditions. The same is expected for H$^{13}$CN, of which the vapour pressure is unknown.
Other isotopic fractionation processes (photolysis and neutral or ionic chemistry) can change the isotopic distribution as a function of altitude. It is the case for the N-bearing molecules in the N$_2$ rich atmosphere of Titan, but the fractionation occurs at very high altitudes, where isotopic ratios are sensitive to magnetospheric electrons \citep{Nosowitz2025,Dobrijevic2018}. In the stratosphere of giant planets, these effects in the C- and N-bearing molecules, such as HCN, are poorly known and more unlikely in the 50\degre S-50\degre N latitude range adopted here, compared to auroral regions. Here, in the absence of observational or theoretical evidence, and to limit the number of parameters to adjust, we did not consider any isotopic fractionation effects as a function of altitude.

Therefore, we derived the H$^{12}$C$^{14}$N/H$^{13}$C$^{14}$N (resp. H$^{12}$C$^{14}$N/H$^{12}$C$^{15}$N) ratios by rescaling the H$^{12}$C$^{14}$N field until the H$^{13}$C$^{14}$N (resp. H$^{12}$C$^{15}$N) line was fitted. The isotopic ratios of carbon and nitrogen in HCN are then simply the following:

\begin{equation*}
    \displaystyle \frac{y_{\mathrm{H}^{12}\mathrm{C}^{14}\mathrm{N}}}{y_{\mathrm{H}^{13}\mathrm{C}^{14}\mathrm{N}}} = 
    \frac{^{12}\mathrm{C}}{^{13}\mathrm{C}} \;\;\;\; \mathrm{and} \;\;\;\; \frac{y_{\mathrm{H}^{12}\mathrm{C}^{14}\mathrm{N}}}{y_{\mathrm{H}^{12}\mathrm{C}^{15}\mathrm{N}}} = \frac{^{14}\mathrm{N}}{^{15}\mathrm{N}} .
\end{equation*}

\section{Results}
\label{part:Results} 

We ran radiative-transfer calculations to find the isotopic ratios that best fit the observations. The results for $^{12}$C/$^{13}$C and $^{14}$N/$^{15}$N are the following:

\begin{align*}
    \frac{^{12}\mathrm{C}}{^{13}\mathrm{C}} &= 73\pm5\\
    & = (0.76-0.87) \times 89.0 \, \mathrm{ (Earth\,ref.)}\\
    & = (0.72-0.83) \times 93.5 \, \mathrm{ (Sun)}\\
    & = (0.69-0.87) \times 92.6 \, \mathrm{ (Jupiter)},
\end{align*}

\begin{align*}
    \frac{^{14}\mathrm{N}}{^{15}\mathrm{N}} &= 245\pm29 \\
    & = (0.80-1.00) \times 272.0 \, \mathrm{(Earth\,ref.)}\\
    & = (0.48-0.63) \times 440.9 \, \mathrm{(Sun)}\\
    & = (0.42-0.70) \times 434.8 \, \mathrm{(Jupiter)}.
\end{align*}

The present isotopic ratios are compared to the reference terrestrial values of 89 for carbon and 272 for nitrogen \citep{IAEA1995,Coplen1992}, to the solar values of $93.5\pm0.7$ and $440.9\pm5.4$ \citep{Lyons2018,Marty2011}, and to the Jovian bulk values from the Galileo entry probe of $92.6\pm4.3$ and $434.8\pm56.7$ \citep{Niemann1998,Owen2001}. In Jupiter, those ratios were measured in CH$_4$ and NH$_3$, respectively, as the major C- and N-bearing species in Jupiter's troposphere. Additional references are found in Tables \ref{tab:RI_C} and \ref{tab:RI_N} for the Jovian bulk. The comparison factors encompass the uncertainties of our derived values in the Jovian exogenic HCN and the uncertainties on the object of comparison (the Earth, Sun, or Jupiter).

We took into account three types of uncertainties that we quadratically added: 1) the natural fitting uncertainty arising from the noise level in the spectrum (at $\pm 1\sigma$); 2) the error arising from the uncertainties on the individual HCN vertical profiles derived by \citet{Cavalie2023b} (see Section \ref{part:HCN_field}); and 3) the uncertainty that arises from the choice of pressure broadening parameters $\gamma$ and $n$ (see Section \ref{part:RT_code}). In order to avoid overestimating the second source of uncertainty when performing the average of the spectra on the 50\degre S-50\degre N latitudinal range, we divided by the square-root of the number of independent pointings in this range ($=303/4$).

For HC$^{15}$N, the major source of uncertainty resides in the noise level (about 70\% of the total error), followed by the effect of the broadening parameters and the uncertainty on the HCN profile (about 15\%). In the case of H$^{13}$CN, the uncertainty arising from the noise level accounts for $\sim$10\%, the rest being due to the broadening parameters and the uncertainty on the HCN profile (about 40\% each). Averaging the spectra from 50\degre S to 50\degre N considerably reduces the impact of the individual HCN vertical profile uncertainties. The uncertainties on the thermal field are already accounted for in the HCN vertical profile uncertainties \citep{Cavalie2023b}. The best fits to the data are shown in Figs. \ref{fig:Resultats_H13CN} and \ref{fig:Resultats_HC15N}. Computations of $\chi^2/N$ values around the line centre give 1.0 and 0.7, respectively, for the H$^{13}$CN and HC$^{15}$N lines.

The above results are representative of the $^{12}$C/$^{13}$C and $^{14}$N/$^{15}$N values in the carbon and nitrogen present in the HCN molecule in Jupiter at the observation date. Computations of the contribution functions, averaged on 50\degre S-50\degre N, indicate that for both isotopologues 60\% of the line emission originates from the 0.03--1.80 mbar range, with a peak at 0.25 mbar. Implications of these results are presented in the next section.

\section{Discussion}
\label{part:Discussion}

Stratospheric HCN is an exogenic species that was formed in extreme conditions from a Jovian air--cometary mixture \citep{Gautier1995,Zahnle1996} and has continuously evolved in the Jovian stratosphere ever since (e.g. \citealt{Moreno2003,Lellouch2006,Cavalie2023b}). Therefore, the interpretation of the isotopic ratios in HCN is particular here; their values years after SL9 can give information on the Jovian--comet mixture at the origin of HCN or on the high pressure or temperature chemistry of the shock, but in principle they can also reflect the history of HCN and its isotopologues in the Jovian stratosphere. These two interpretations, i.e. origin and evolution, are discussed in this section.

In this context, we now have two data points for the $^{12}$C/$^{13}$C and $^{14}$N/$^{15}$N ratios in HCN, one from 1998 observations \citep{Matthews2002}, and one from 2017 data. No observations of the heavier isotopologues of HCN (or of any other exogenic species containing carbon or nitrogen) were made right after the collisions; hence, we do not have any direct constraints on the isotopic composition at that time and we can only rely on later measurements. A comparison of the 1998 and 2017 results shows a contradictory behaviour; i.e. while the Matthews et al. measurements suggest an atypically strong depletion in both $^{13}$C and $^{15}$N relative to the Jovian bulk, our new estimates instead indicate an enrichment compared to Jovian values. In what follows, we first discuss the validity of the 1998 results in Section \ref{dis:1998results}. We explain why we eventually chose to discard them. We then give our interpretation of the 2017 results alone in Section \ref{dis:2017results}.

\subsection{Implications of the 1998 measurements}
\label{dis:1998results}
The first observations of H$^{13}$CN and HC$^{15}$N were taken in September 1998, four years after the SL9 impacts, by \citet{Matthews2002}. They found the following ranges of values:

\begin{align*}
    \frac{^{12}\mathrm{C}}{^{13}\mathrm{C}} &= 176-409 \\
    & = (2.0-4.6) \times 89 \, \mathrm{ (Earth\,ref.)}\\
    & = (1.9-4.4) \times 93.5 \, \mathrm{ (Sun)}\\
    & = (1.8-4.6) \times 92.6 \, \mathrm{ (Jupiter),}
\end{align*}

\begin{align*}
    \frac{^{14}\mathrm{N}}{^{15}\mathrm{N}} &= 1150-4500\\
    & = (4.3-16.7) \times 272 \, \mathrm{(Earth\,ref.)}\\
    & = (2.5-10.3) \times 440.9 \, \mathrm{(Sun)}\\
    & = (1.8-11.2) \times 434.8 \, \mathrm{(Jupiter).}\\
\end{align*}

Their results are compared to the Earth reference and to the solar and Jovian bulks, in the same way as in Section \ref{part:Results}. Figs. \ref{fig:RI_C} and \ref{fig:RI_N} present determinations of the carbon and nitrogen ratios in various Solar System objects (more details are found in Tables \ref{tab:RI_C} to \ref{tab:RI_N} and references therein). The Jovian atmosphere is well known to present solar values both for carbon and nitrogen (see the most accurate results from the in situ measurements of the Galileo entry probe (\citealt{Niemann1998,Owen2001}); a Jovian source was consequently ruled out. \citet{Matthews2002} interpreted the ratios to originate from an unusual cometary composition or an unknown isotopic fractionation process during the different phases of the impacts. We discuss the two hypotheses below and also give other interpretations.

\subsubsection{An unknown isotopic fractionation during the impacts}
Firstly, one possibility is that a fractionation mechanism in the shocks could have preferentially enriched HCN in the lighter isotopes ($^{12}$C and $^{14}$N), probably through interactions with the dust, as suggested by \citet{Moreno2003}. However, this hypothesis would need to be confirmed by laboratory experiments and/or plume modelling. Such work should explain why C, N, and S seem to be affected by this depletion process similarly, even if their isotopic ratios are usually very different. Until then, this possibility is still open.

\subsubsection{An unusual comet composition}
More than 20 years since \citet{Matthews2002}, we now have access to tens of derivations in different comets and in several molecules. Among these, one can note the in situ data of the comet 67P/Churyumov--Gerasimenko in various molecules. All the measurements show Earth-like values for carbon, and the nitrogen ratio is systematically enriched in $^{15}$N (as measured in several N-bearing compounds) with respect to the terrestrial value with factors up to five, and this is, on average, with a factor of two (see Figs. \ref{fig:RI_C} and \ref{fig:RI_N}). For the $^{12}$C/$^{13}$C ratio, only \citet{Biver2016} nominally found a depletion in the comet C/2012 F6 (Lemmon) with respect to the terrestrial value, but their error bars are very large (124$\pm$64) compared to many other estimates, and they still encompass the terrestrial value. For the $^{14}$N/$^{15}$N ratio, even the highest cometary values (e.g. $205\pm70$ in HCN in comet C/1995 01 (Hale--Bopp; \citealt{Hily-Blant2017}) are at least more enriched in $^{15}$N by a factor of five compared to those of \citet{Matthews2002}. An unusual cometary source for the carbon and nitrogen content in HCN, as proposed by \citet{Matthews2002}, is thus highly improbable with regard to all these comet measurements.

\subsubsection{A temporal evolution of HCN between 1994 and 1998}
Another potential explanation may derive from a temporal evolution of the total mass of HCN after the comet impacts of 1994, altering the isotopic composition of HCN. Continuous observations after the collisions indicated that the total mass of HCN increased by a factor up to six \citep{Moreno2003} ten months after SL9, compared to the initial HCN mass derived by \citet{Bezard1997a}. The same amount was still observed in September 1998 by \citet{Moreno2003} and was consistent with the \textit{Cassini}/CIRS observations at the end of 2000 (\citealt{Lellouch2006} indicated an increase by a factor of $5.5^{+5.0}_{-2.5}$). As discussed extensively in \citet{Lellouch2006}, the main explanation of the HCN enhancement is the conversion of Jovian NH$_3$ \citep{Moses1996book} which was observed in the upper stratosphere after SL9 \citep{Griffith1997}. 

This additional production of HCN from NH$_3$ should ultimately have modified the isotopic composition of the initial SL9-produced HCN. With very simple considerations, we can give an estimation of the isotopic ratios of the initial SL9-produced HCN, noted  $\mathrm{IR}_{i}$. The values of \citet{Matthews2002}, denoted $\mathrm{IR}_{f}$, are taken as the isotopic ratios of the total amount of HCN in 1998 (after the NH$_3$ conversion). The additional NH$_3$-produced HCN should present the typical ratios of the Jovian NH$_3$ and CH$_4$ denoted $\mathrm{IR}_{a}$, taken here as the values measured by the Galileo entry probe (see Tables \ref{tab:RI_C} and \ref{tab:RI_N}). The isotopic ratios of the initial SL9-produced HCN are written as follows (for carbon or nitrogen):

\begin{equation*}
    \mathrm{IR}_{i} = \left( \frac{n}{\mathrm{IR}_{f}} + \frac{1-n}{\mathrm{IR}_{a}}\right)^{-1},
\end{equation*}

\noindent where $n$ is the enrichment factor of HCN between the initial SL9-produced amount and the total amount of HCN several months after the impacts. The above formula is only valid for $n \geq 1$, because no loss factor of HCN is considered. Values of $n$ lie between 0.6 and six following the results of \citet{Moreno2003} and between three and 10.5 according to \citet{Lellouch2006}. As a consequence, we considered $n$ ranges between three and six. 

Considering the uncertainties on $\mathrm{IR}_{f}$ and $\mathrm{IR}_{a}$, $\mathrm{IR}_{i}$ presents only negative (i.e. non-physical) values for $n>1.9$ for carbon and for $n>1.4$ for nitrogen. Positive values can exist only if $n$ is close to one; in this case, the initial ratios become even higher than those of \citet{Matthews2002}, rendering them even more incompatible with values measured in comets. In addition, such values for $n$ are incompatible with the range resulting from \citet{Moreno2003} and \citet{Lellouch2006}. In conclusion, the changes in the HCN abundance in the years after the impacts cannot explain the abnormal isotopic ratios reported by \citet{Matthews2002}.

\subsubsection{A possible modelling issue}
\citet{Matthews2002} analysed HCN (J=4--3) and CS (J=7--6) spectra obtained at the equatorial limb and at the planet's disc centre with the James Clerk Maxwell Telescope (JCMT). The half power beam width (HPBW) was 13\arcsec.5, and the angular diameter of Jupiter was 50\arcsec, resulting in a covered latitudinal range of 10\degre S-10\degre N. At the time of their observations, HCN had already spread out latitudinally and was present up to northern mid-latitudes with a mixing ratio of about $10^{-8}$ \citep{Moreno2003}.

Independent temperature information was unavailable for the modelling of their observations. As a consequence, \citet{Matthews2002} simultaneously derived the temperature vertical profile and the HCN and CS abundance vertical profiles directly from the HCN and CS lines using the different observation geometries (limb vs. nadir). They used simple constant vertical profiles for P$_0$<0.2 mbar both for the temperature and the HCN abundance. Because the H$^{12}$C$^{14}$N line was optically thick ($\tau \sim$ 10 at the line centre), the H$^{12}$C$^{14}$N abundance that is derived is strongly tied to the retrieved temperature. In addition, the P$_0$ cut-off pressure derived from the line width also shows some dependency on the temperature. Thus, several solutions can be found simultaneously for the temperature, the HCN abundance, and P$_0$. To illustrate this point, \citet{Matthews2002} found an HCN abundance ranging from (3.0-4.5)$\times 10^{-8}$ for a temperature of $164\pm2$ K at P$_0$<0.2 mbar, while a reanalysis in \citet{Moreno2003} indicated (0.8-2.5)$\times 10^{-8}$ for a temperature of $167\pm1$ K at P$_0$<0.3 mbar for the same dataset.
Our conclusion is that the abundance of H$^{12}$C$^{14}$N used to compute the isotopic ratios might not be well constrained, and this immediately impacts the isotopic ratios values. As a comparison, the CS line was less optically thick, and the subsequent sulphur ratio was closer to the terrestrial value, but still depleted in $^{34}$S.

Contrary to H$^{12}$C$^{14}$N, the H$^{13}$CN and HC$^{15}$N lines were optically thin in 1998, and the derived abundances are more reliable. Therefore, a better estimator of the isotopic composition in HCN can reside in the H$^{13}$CN/HC$^{15}$N ratio. As temperature should affect the determination of the H$^{13}$CN and HC$^{15}$N abundances in a similar way, the H$^{13}$CN/HC$^{15}$N ratio should not be too temperature dependent. \citet{Matthews2002} found volume-mixing ratios of $\left(1.4\pm0.3\right)\times 10^{-10}$ and $\left(1.8\pm0.8\right)\times 10^{-11}$, respectively, for H$^{13}$CN and HC$^{15}$N. From these values and their uncertainties, acceptable ratios are in the H$^{13}$CN/HC$^{15}$N=(2.7--12.9) range. As a comparison, the terrestrial and Jovian ratios give, respectively, 3.1 using the Earth references and (3.9--5.5) using the values of the Galileo entry probe. Our results lead to H$^{13}$CN/HC$^{15}$N=(2.8-4.0). Thus, when not considering the H$^{12}$C$^{14}$N abundance, they are now compatible with those of \citet{Matthews2002}, within their large error bars.\\

For all the above reasons, we deem the isotopic ratios of \citet{Matthews2002} not to be sufficiently reliable, and useful information can only be found in their $^{13}$C/$^{15}$N ratio. In the next section, we disregard them and only consider our new estimates to propose new scenarios on the origin and evolution of HCN.

\subsection{Implications of the 2017 results}
\label{dis:2017results}

\begin{figure*}[!ht]
\centering
\includegraphics[width=\textwidth]{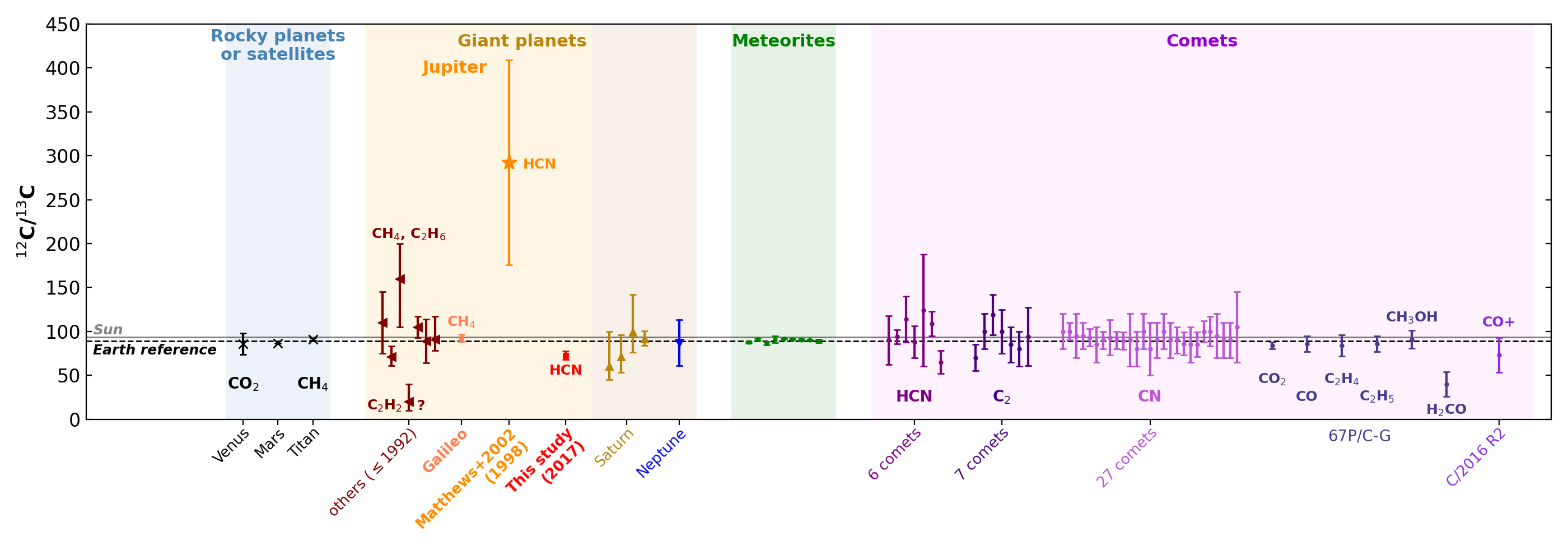}
\caption{Review of $^{12}$C/$^{13}$C ratio in the various Solar System objects. Detailed information on the instrument, the observation date, and the references are reported in Tables \ref{tab:RI_C} (giant planets only) and \ref{tab:RI_C_2} (the other Solar System determinations) and references therein.}
\label{fig:RI_C}
\end{figure*}

\begin{figure*}[!ht]
\centering
\includegraphics[width=\textwidth]{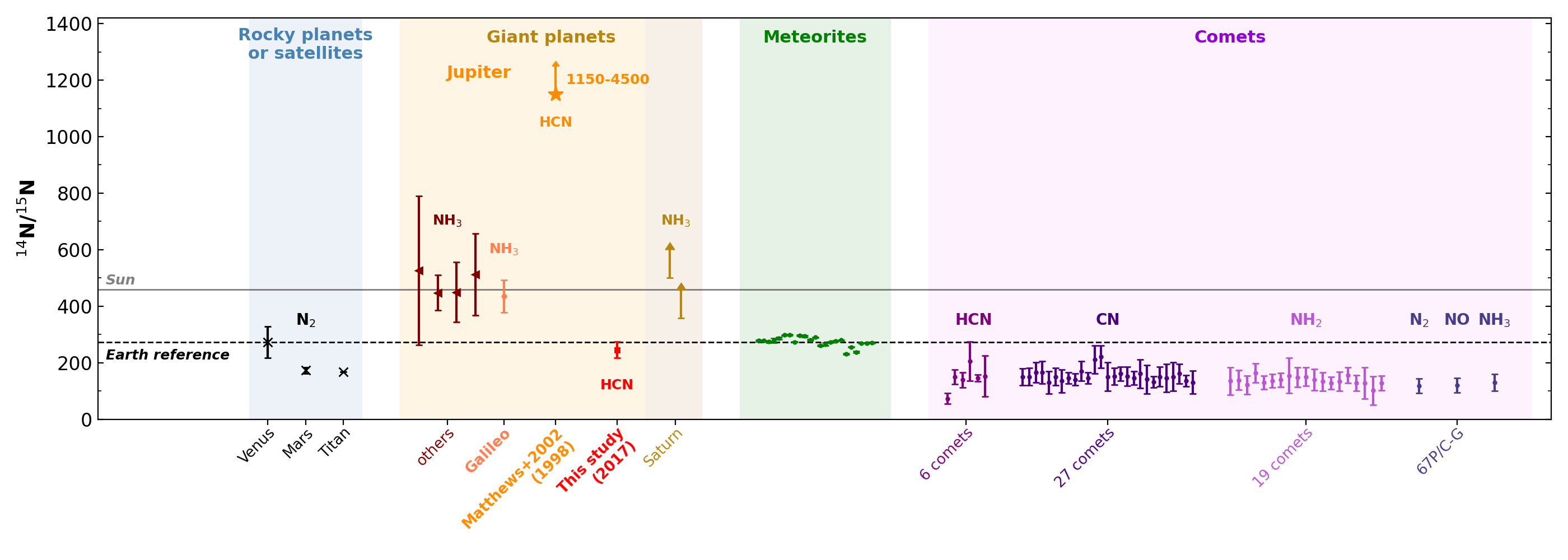}
\caption{Review of $^{14}$N/$^{15}$N ratio in the various Solar System objects. Detailed information on the instrument, the observation date, and the references are reported in Table \ref{tab:RI_N} and references therein.}
\label{fig:RI_N}
\end{figure*}

We derived new carbon and nitrogen isotopic ratios in HCN 23 years after the SL9 impacts. We find that HCN in 2017 is enriched in $^{13}$C and $^{15}$N compared to the solar--Jovian bulk values. The ranges of values defined by the uncertainties at $1\sigma$ are incompatible with the ratios in Jupiter for both ratios. For nitrogen, the $^{14}$N/$^{15}$N ratio still encompasses the more $^{15}$N-enriched terrestrial value, but on the edge of the error bar. The nitrogen ratio appears to be at an intermediate value between the Jovian bulk and typical cometary values that are $^{15}$N-rich (see Fig. \ref{fig:RI_N}). For carbon, the comparison is less diagnostic as all the Solar System objects exhibit very similar values (see Fig. \ref{fig:RI_C}), but a value of $73\pm5$ is still unusual in the Solar System. Enrichments in $^{13}$C with respect to the terrestrial reference are observed in some comets within the error bars. Therefore, the enrichments seen here could constitute the direct signature of the comet. 

However, during the 23 years separating our measurements and the impacts of SL9, we recall that HCN evolved in the Jovian atmosphere by spreading out horizontally and vertically \citep{Moreno2003,Lellouch2006,Cavalie2023b}, but also chemically. Several chemical processes may have taken place or may still be taking place. 1) In the months following SL9, additional HCN was produced from stratospheric NH$_3$ conversion \citep{Moreno2003,Lellouch2006}; 2) monitoring years later in 2017 by \citet{Cavalie2023b} suggested that the total quantity of HCN decreased by a factor of $5.0\pm3.0$ compared to the 1995--1998 data of \citet{Moreno2003} or $12.0\pm3.5$ compared to the 2000 values of \citet{Lellouch2006}. This decrease of the HCN mass over time is mainly caused by auroral losses through adsorption on aerosols; 3) in addition, several isotopic fractionation processes could have occurred between 1994 and 2017 and could have altered the isotopic composition (neutral or ionic chemistry in the gas phase, chemistry on aerosols, photolysis, etc.). In the N$_2$-rich atmosphere of Titan, isotopic fractionation in N-bearing molecules exists and occurs as a function of altitude and depending on the molecule \citep{Nosowitz2025,Dobrijevic2018}. However, little is known about the fractionation rates and the kinetics of these potential reactions with HCN at the $\sim$0.1 mbar level in the Jovian atmosphere. 

In what follows, for the sake of simplicity, we give an estimation of the origin of the elements in HCN, without any isotopic fractionation mechanism over time. Contrary to carbon, the nitrogen isotopic ratio shows very distinct values depending on the objects, and it is therefore a more reliable tracer of the origin. We can thus estimate the nitrogen contribution from the comet to HCN.

Following the method used in \citet{Matthews2002}, the total number of an atom, X ($^{14}$N or $^{15}$N), is expressed as $\mathrm{N}^\mathrm{X}_{\mathrm{tot}} = \mathrm{N}^\mathrm{X}_{\mathrm{SL9}} + \mathrm{N}^\mathrm{X}_{\mathrm{J}}$, where $\mathrm{N}^\mathrm{X}_{\mathrm{SL9}}$ is the number of the X atom coming from the comet, and similarly $\mathrm{N}^\mathrm{X}_{\mathrm{J}}$ from Jupiter. The nitrogen isotopic ratios in the mixture, in SL9, and in Jupiter are written as $\mathrm{IR}_{\mathrm{tot,SL9,J}} = \mathrm{N}^{^{14}\mathrm{N}}_{\mathrm{tot,SL9,J}}/\mathrm{N}^{^{15}\mathrm{N}}_{\mathrm{tot,SL9,J}}$. The fraction of Jovian nitrogen in the mixture, $f=\frac{\mathrm{N}^{^{14}\mathrm{N}}_{\mathrm{J}}}{\mathrm{N}^{^{14}\mathrm{N}}_{\mathrm{tot}}}$, is given by

\begin{equation*}
    f = \frac{\mathrm{IR}_{\mathrm{J}}\times(\mathrm{IR}_{\mathrm{SL9}}-\mathrm{IR}_{\mathrm{tot}})}{\mathrm{IR}_{\mathrm{tot}}\times(\mathrm{IR}_{\mathrm{SL9}}-\mathrm{IR}_{\mathrm{J}})},
\end{equation*}

\noindent where $\mathrm{IR}_{\mathrm{tot}}$ is the value of the mixture that we derived here, $\mathrm{IR}_{\mathrm{J}}$ is well known from Galileo entry probe measurements \citep{Owen2001}, and $\mathrm{IR}_{\mathrm{SL9}}$ is taken as the weighted average of the $\sim$ 50 measurements presented in Fig. \ref{fig:RI_N} and referenced in Table \ref{tab:RI_N}, leading to $\left(140\pm4\right)$.

We find the fraction of cometary nitrogen to be $\sim$(40$\pm$20)\% of the total amount of nitrogen in HCN, taking into account the uncertainties on each ratio. The fraction of Jovian nitrogen is thus (60$\pm$20)\%. The latter contains all the Jovian contributions, including the one resulting from the conversion of stratospheric NH$_3$ in the months that followed the impacts. Removing the latter from the total Jovian contribution depends on the HCN increase factor, $n$, which is not tightly constrained (see \citealt{Moreno2003,Lellouch2006}, and Section \ref{dis:1998results}). For example, with $n=3$ or $n=5$, about (50--80)\% or (60--90)\% (respectively) of the nitrogen in the initially SL9-produced HCN would be cometary. These results then suggest at least an equal cometary and Jovian origin for nitrogen in the SL9-produced HCN. As discussed in \citet{Lellouch1996book}, a predominant cometary origin for nitrogen is possible in the exogenic N-bearing molecules, which is consistent with our estimations.\\

New HCN, H$^{13}$CN, and HC$^{15}$N data are needed to monitor the temporal evolution of the isotopic ratios in HCN. A variation of the ratios after our 2017 measurements would definitively point towards a fractionation mechanism as a function of time, induced by various chemical reactions. If such a variation is seen, better constraints will be set on the poorly known chemistry between HCN, its isotopologues, and the rest of the Jovian atmosphere. If, within the error bars, no temporal evolution of the ratios is observed, it would either suggest that a chemical equilibrium has already been reached or that isotopic fractionation is negligible on a timescale of a decade. In the last case, the isotopic ratios could then directly represent the initial mixture at the origin of the exogenic molecules formed during the SL9 impacts. In the meantime, photochemical models are needed here to confront the enrichments we derived to potential fractionation processes and to understand how these processes alter the isotopic composition over time.

However, as time is passing by, the abundance of the SL9-produced HCN is expected to continue its decrease at mid-latitudes from photolysis, vertical transport, and horizontal transport/destruction towards the poles, making it more difficult to obtain good SNR measurements. The JUpiter ICy moons Explorer (Juice) mission will arrive at Jupiter in 2031, and monitoring of the isotopic ratios in HCN will be of great importance to understand the isotopologues' evolution and chemistry \citep{Fletcher2023}. It will be especially interesting to look at the isotopologues of other well-known molecules to try to complete the picture of fractionation processes as a function of the species.

\section{Conclusion}
\label{part:Conclusion} 

We present ALMA observations of Jupiter of H$^{13}$CN and HC$^{15}$N lines at 345.3398 and 344.2003 GHz, respectively, taken in 2017. To increase the SNR, we co-added limb spectra on the latitudinal range 50\degre S-50\degre N after correcting for the planet-induced Doppler shift of the lines observed at different latitudes on the planet limb. With a radiative-transfer model, we rescaled the HCN vertical profiles retrieved from the same ALMA dataset \citep{Cavalie2023b} to fit the isotopologue lines. The $^{12}$C/$^{13}$C and $^{14}$N/$^{15}$N isotopic ratios deduced from the fitting are $73\pm5$ and $245\pm29$. The values indicate that HCN is enriched in the heavier isotopes compared to the Jovian bulk values measured in CH$_4$ and NH$_3$. This behaviour is at odds with the 1998 measurements of \citet{Matthews2002} that suggested a very unusual depletion. Because of limitations in the retrieval of the H$^{12}$C$^{14}$N abundance, we chose to disregard their results when discussing our new measurements.

Using only our new derivations, we conclude that the enrichments we see here in both $^{13}$C and $^{15}$N can result from two, not mutually exclusive, scenarios: 1) the signature of the comet contribution to the formation of the Jovian stratospheric HCN and/or 2) the chemical evolution of HCN in the Jovian atmosphere. In the absence of models taking into account the different fractionation mechanisms for HCN, we can only give an estimation of the origin of the material in HCN, without considering fractionation over time. Since the carbon isotopic ratio does not seem to present any fractionation in Solar System objects compared to the terrestrial/solar values, it is more difficult to conclude on the cometary contribution to carbon, especially since our value is inconsistent with the standard terrestrial and solar values. On the contrary, comets are easily traceable by their highly $^{15}$N-enriched ratios compared to the Jovian value close to solar. With the knowledge of the bulk isotopic composition of Jupiter, the dozens of derivations in comets, our new measurement, and by taking into account the additional production of HCN from NH$_3$ in the months after the SL9 impacts, we estimate the cometary contribution to the nitrogen composing the initial SL9-produced HCN to be at least 50\%.

Monitoring the isotopic ratios in HCN over time is essential to constrain the chemical evolution of the minor isotopologues and infer the possible fractionation processes. Future observations with the Juice spacecraft will be especially needed in that respect. Observations should be confronted with photochemical models that incorporate the isotopologue chemistry to give an idea of the strength and the timescale of the fractionation processes. However, one should keep in mind that the HCN abundance at mid-latitudes decreases over time because of transport and photolysis. Thus, achieving high SNRs of the minor isotopologue lines will become increasingly challenging, limiting our ability to determine the isotopic composition of HCN.

\section*{Acknowledgements}
C. Lefour and T. Cavalié acknowledge funding from the Centre National d'Études Spatiales (CNES) and the Programme National de Planétologie (PNP) of CNRS/INSU. \\
This Paper makes use of the following ALMA data: ADS/JAO.ALMA\#2016.1.01235.S. ALMA is a partnership of ESO (representing its member states), NSF (USA) and NINS (Japan), together with NRC (Canada), MOST and ASIAA (Taiwan), and KASI (Republic of Korea), in cooperation with the Republic of Chile. The Joint ALMA Observatory is operated by ESO, AUI/NRAO and NAOJ.

\newpage
\begin{appendix}

\section{Carbon and nitrogen isotopic ratios in the Solar System}

Tables \ref{tab:RI_C} and \ref{tab:RI_C_2} summarise $^{12}$C/$^{13}$C determinations in various Solar System objects (giant planets in Table \ref{tab:RI_C} and other objects in Table \ref{tab:RI_C_2}). Table \ref{tab:RI_N} presents similar data but for the $^{14}$N/$^{15}$N ratio. Details on the object, molecule, instrument, observation date, sensitive region, and the references from which we extracted the data are added. Figs. \ref{fig:RI_C} and \ref{fig:RI_N} display these values. In the case of comets and meteorites, as many measurements were performed, we do not detail them in the tables below, but values can be found in the references and are also illustrated in Figs. \ref{fig:RI_C} and \ref{fig:RI_N}.

\begin{sidewaystable*}
\centering
\caption{Review of the $^{12}$C/$^{13}$C ratio in the atmosphere of giant planets and various Solar System objects.}
\label{tab:RI_C}

\begin{tabular}{ccccccc}
\specialrule{.2em}{.1em}{.1em} 

Object & Molecule & Technical information & Date & $^{12}$C/$^{13}$C & Sensitive region & References \\ \specialrule{.2em}{.1em}{.1em} 

 \multicolumn{7}{c}{Giant planets} \\ \hline \hline
 Jupiter & HCN        & ALMA - 345 GHz                          & 2017      & $73\pm5$  & 0.1-1 mbar    & 1 \\ \hline

 Jupiter & HCN        & JCMT - 354 GHz                          & 1998      & 176-409            & <0.2 mbar     & 2 \\ \hline

 Jupiter & CH$_4$     & Galileo/GPMS (\textit{in situ})         & 1995      & $92.6 \pm 4.3$         & 0.5-3.8 bar   & 3 \\ \hline

 Jupiter & C$_2$H$_6$ & McMath-Pierce Telescope - 822 cm$^{-1}$ & 1995      & $91^{+26}_{-13}$       & 5 mbar        & 4 \\ \hline

 Jupiter & CH$_4$     & CFHT - 2450-2600 cm$^{-1}$              & 1992      & $89 \pm 25$            & ?             & 5 \\ \hline

 Jupiter & C$_2$H$_6$ & KPNO - 822 cm$^{-1}$                    & 1987      & $105 \pm 12$           & 10 mbar       & 6, 4 \\ \hline

 Jupiter & C$_2$H$_2$ & IRTF - 755 cm$^{-1}$                    & 1984      & \makecell[c]{$20^{+20}_{-10}$\\qualified as "uncertain"} & <20 mbar & 7 \\ \hline

 Jupiter & CH$_4$     & Voyager 1 - 1300 cm$^{-1}$              & 1979      & $160^{+40}_{-55}$      & stratosphere  & 8 \\ \hline

 Jupiter & CH$_4$     & Mont Palomar Observatory - 9050 cm$^{-1}$ & 1974    & $71^{+12}_{-10}$       & 600 mbar      & 9, 4 \\ \hline

 Jupiter & CH$_4$     & McDonald Observatory - 9050 cm$^{-1}$     & 1972    & $110 \pm 35$           & 700-1100 mbar & 10 \\ \specialrule{.2em}{.1em}{.1em}

 Saturn  & CH$_4$     & Cassini/CIRS - 1100-1400 $^{-1}$        & 2004-2007 & $91.8^{+8.4}_{-7.8}$   & 1.0-6.1 mbar  & 11 \\ \hline      

 Saturn  & C$_2$H$_6$ & McMath-Pierce Telescope - 822 cm$^{-1}$ & 1995      & $99^{+43}_{-23}$       & 5 mbar        & 4 \\ \hline

 Saturn  & CH$_4$     & Mont Palomar Observatory - 9050 cm$^{-1}$ & 1974    & $71^{+25}_{-18}$       & 600 mbar      & 9 \\ \hline

 Saturn  & CH$_4$     & OHP - 9050 cm$^{-1}$                      & 1972    & $60^{+40}_{-15}$       & 400-900 mbar  & 12 \\ \specialrule{.2em}{.1em}{.1em}

 Neptune & C$_2$H$_6$ & IRTF - 822 cm$^{-1}$                    & 1990      & $87 \pm 26$            & 0.1-1 mbar    & 13 \\ \specialrule{.2em}{.1em}{.1em}

\end{tabular}

\vspace{2mm}
\noindent\textbf{Notes.} First and second columns list the name of the object and the molecule in which the isotopic ratio was measured. Third and fourth columns give information on the instrument and the date of the observation. The isotopic ratios are listed in the fifth column. For planets only, the region in the atmosphere that is probed by the observations is given in the sixth column, if known. Values extracted from the references below are presented in Fig. \ref{fig:RI_C}.

\vspace{2mm}
\noindent\textbf{References.} 
(1) this study; 
(2) \citet{Matthews2002}; 
(3) \citet{Niemann1998}; 
(4) \citet{Sada1996}; 
(5) \citet{Marten1994}; 
(6) \citet{Wiedemann1991}, revised in \citet{Sada1996}; 
(7) \citet{Drossart1985b}; 
(8) \citet{Courtin1983}; 
(9) \citet{Combes1977} (and references therein), revised in \citet{Sada1996}; 
(10) \citet{Fox1972}; 
(11) \citet{Fletcher2009a}; 
(12) \citet{Combes1975}; 
(13) \citet{Orton1992}, revised in \citet{Sada1996}.

\end{sidewaystable*}

\begin{sidewaystable*}
\centering
\caption{Table \ref{tab:RI_C} continued.}
\label{tab:RI_C_2}

\begin{tabular}{ccccccc}
\specialrule{.2em}{.1em}{.1em} 

Object & Molecule & Technical information & Date & $^{12}$C/$^{13}$C & Sensitive region & References \\ \specialrule{.2em}{.1em}{.1em} 
\multicolumn{7}{c}{Other Solar System objects} \\ \hline \hline
 Sun     & CO     & ATMOS FTS  & 1995           & $93.5 \pm 0.7$ & photosphere & 1 \\ \hline
 
 Earth   & C      & Reference  & -              & 89             & \makecell[c]{Pee Dee \\ Belemnite (PDB)} & 2 \\ \hline

 Venus   & CO$_2$ & OHP        & 1985    & $86 \pm 12$       & atmosphere       & 3 \\ \hline

 Mars    & C      & Landers, meteorites, GB telescopes & -    & \makecell[c]{$86.5 \pm 0.2$ (mean) \\ 81-93 (range)}    & -       & 4 \\ \hline

 Titan   & CH$_4$ & Huygens entry probe (\textit{in situ}) & 2005 & $91.1 \pm 1.4$    & <127 km & 5 \\ \specialrule{.2em}{.1em}{.1em} 

 Meteorites      & C      & -   & -    & \makecell[c]{$91.4 \pm 0.1$ (mean) \\ 84.5-95.0 (range)} & -          & 6, 7 \\ \specialrule{.2em}{.1em}{.1em}

 6 comets  & HCN  & -         & - & \makecell[c]{$91.5 \pm 5.5$ (mean $^{(a)}$) \\ 52-188 (range $^{(b)}$)}& coma   & 8, 9, 10, 11 \\ \hline

 7 comets  & C$_2$& -         & - & \makecell[c]{$88.4 \pm 7.8$ (mean) \\ 55-142 (range)}    & coma   & 12 \\ \hline              
 
 27 comets & CN   & -         & - & \makecell[c]{$92.2 \pm 3.0$ (mean) \\ 50-145 (range)}    & coma   & 12, 13, 14, 15, 16 \\ \hline                     

 Comet 67P/ & \makecell[c]{CO$_2$ \\ CO \\ C$_2$H$_4$ \\ C$_2$H$_5$ \\ CH$_3$OH \\ H$_2$CO}    & Rosetta/ROSINA (\textit{in situ})  & 2015 & \makecell[c]{$84 \pm 4$ \\ $86 \pm 9$ \\ $84 \pm 12$ \\ $86 \pm 9$ \\ $91 \pm 10$ \\ $40 \pm 14$}  & coma & 17, 18, 19 \\ \hline

 Comet C/2016 R2 & CO$^+$ & VLT & 2018 & $73 \pm 20$                      & coma       & 20 \\ \specialrule{.2em}{.1em}{.1em} 

\end{tabular}

\vspace{2mm}
\noindent\textbf{Notes.} Description of columns is the same as in Table \ref{tab:RI_C}. Values extracted from the references below are presented in Fig. \ref{fig:RI_C}.\\
(a) The mean values are averages weighted by the individual uncertainties. The uncertainties correspond to the uncertainty on the weighted average. \\
(b) This interval show the amplitude of values covered within the error bars of the different measurements.

\vspace{2mm}
\noindent\textbf{References.} 
(1) \citet{Lyons2018}; 
(2) \citet{IAEA1995}; 
(3) \citet{Bezard1987}; 
(4) \citet{Webster2013} and references therein; 
(5) \citet{Niemann2010}; 
(6) \citet{Woods2009}; 
(7) \citet{Woods2009DB}; 
(8) \citet{Cordiner2019}; 
(9) \citet{Cordiner2024}; 
(10) \citet{Biver2016}; 
(11) \citet{Bockelee2008}; 
(12) \citet{Bockelee2015} and references therein; 
(13) \citet{Moulane2020}; 
(14) \citet{Moulane2023}; 
(15) \citet{Manfroid2009}; 
(16) \citet{Wyckoff2000}; 
(17) \citet{Hassig2017} (CO$_2$); 
(18) \citet{Rubin2017} (CO, C$_2$H$_4$, C$_2$H$_5$); 
(19) \citet{Altwegg2020} (CH$_3$OH, H$_2$CO); 
(20) \citet{Rousselot2024}.

\end{sidewaystable*}

\begin{sidewaystable*}
\centering
\caption{Review of the $^{14}$N/$^{15}$N ratio in the atmosphere of giant planets and various Solar System objects.}
\label{tab:RI_N}

\begin{tabular}{ccccccc}
\specialrule{.2em}{.1em}{.1em} 

Object & Molecule & Technical information & Date & $^{14}$N/$^{15}$N & Sensitive region & References \\ \specialrule{.2em}{.1em}{.1em} 

\multicolumn{7}{c}{Giant planets} \\ \hline \hline
 Jupiter & HCN    & ALMA - 344 GHz                                & 2017          & $245\pm29$     & 0.1-1 mbar     & 1 \\ \hline
 
 Jupiter & NH$_3$ & IRTF/TEXES - 900-960 cm$^{-1}$  & 2013          & $513 \pm 145$                      & troposphere    & 2 \\ \hline

 Jupiter & NH$_3$ & Cassini/CIRS - 900-940 cm$^{-1}$ & 2000          & $450\pm106$                 & 0.5 bar        & 3 \\ \hline

 Jupiter & NH$_3$ & Cassini/CIRS - 863-903 cm$^{-1}$ & 2000          & $448\pm62$                 & 0.5 bar        & 4 \\ \hline
 
 Jupiter & HCN    & JCMT - 344 GHz                               & 1998           & 1150-4500                     & <0.2 mbar      & 5 \\ \hline

 Jupiter & NH$_3$ & ISO/SWS - 10 $\mu$m                   & 1996-1997 & $526\pm263$           & 0.4 bar        & 6 \\ \hline

 Jupiter & NH$_3$ & Galileo/GPMS (\textit{in situ})         & 1995          & $434.8\pm56.7$                   & 0.9-2.9 bar    & 7 \\

 \specialrule{.2em}{.1em}{.1em}

  Saturn  & NH$_3$ & IRTF/TEXES - 900-960 cm$^{-1}$   & 2012-2013 & \makecell[c]{>500 (900 cm$^{-1}$) \\ >357 (960 cm$^{-1}$)} & troposphere & 2 \\ \specialrule{.2em}{.1em}{.1em}

\multicolumn{7}{c}{Other Solar System objects} \\ \hline \hline
 Bulk Sun          & N     & Genesis solar wind samples return & 2001-2004 & $440.9\pm5.4$   & solar wind    & 8 \\ \hline

 Earth              & N$_2$ & IUPAC reference                           & -                  & 272                    & atmosphere & 9 \\ \hline

 Venus              & N$_2$ & Pioneer Venus probe               & 1978       & $273 \pm 56$               & atmosphere & 10 \\ \hline

 Mars               & N$_2$ & MSL                               & 2013       & $173 \pm 11$               & atmosphere & 11 \\ \hline

 Titan              & N$_2$ & Huygens entry probe (\textit{in situ}) & 2005  & $167.7 \pm 0.6$         & <144 km & 12 \\ \specialrule{.2em}{.1em}{.1em}

 Meteorites         & N     &       -                            & -         & \makecell[c]{$278.6 \pm 0.2$ (mean)\\230-300 (range)}         & -       & 13 \\ \specialrule{.2em}{.1em}{.1em}

 Comet 67P/ & \makecell[c]{N$_2$\\NO\\NH$_3$} & Rosetta/ROSINA (\textit{in situ}) & 2015 & \makecell[c]{$118 \pm 25$ \\ $120 \pm 25$ \\ $130 \pm 30$} & coma & 14 \\ \hline

 \multirow{2}{*}{6 comets}            & \multirow{2}{*}{HCN}   & \multirow{2}{*}{-} & \multirow{2}{*}{-} & $130 \pm 9$ (mean)  & \multirow{6}{*}{coma} & \multirow{6}{*}{15, 16, 17, 18} \\
                                      &                        &                    &                    & 54-275 (range)      &                       &  \\ 
                                      
 \multirow{2}{*}{27 comets/fragments} & \multirow{2}{*}{CN}    & \multirow{2}{*}{-} & \multirow{2}{*}{-} & $149 \pm 6$ (mean)  &                       & \\ 
                                      &                        &                    &                    & 90-261 (range)      &                       &  \\
                                      
 \multirow{2}{*}{19 comets/fragments} & \multirow{2}{*}{NH$_2$}& \multirow{2}{*}{-} & \multirow{2}{*}{-} & $136 \pm 7$ (mean)  &                       &   \\ 
                                      &                        &                    &                    & 51-217 (range)      &                       &   \\ \specialrule{.2em}{.1em}{.1em}

\end{tabular}

\vspace{2mm}
\noindent\textbf{Notes.} Description of columns is the same as in Table \ref{tab:RI_C}. Values extracted from the references below are presented in Fig. \ref{fig:RI_N}.

\vspace{2mm}
\noindent\textbf{References.} 
(1) this study; 
(2) \citet{Fletcher2014}; 
(3) \citet{Fouchet2004}; 
(4) \citet{Abbas2004}; 
(5) \citet{Matthews2002}; 
(6) \citet{Fouchet2000}; 
(7) \citet{Owen2001}; 
(8) \citet{Marty2011} ; 
(9) \citet{Coplen1992}; 
(10) \citet{Hoffman1979}; 
(11) \citet{Wong2013}; 
(12) \citet{Niemann2010}; 
(13) \citet{Grewal2022}; 
(14) \citet{Altwegg2019} and references therein; 
(15) \citet{Cordiner2024}; 
(16) \citet{Biver2021}; 
(17) \citet{Moulane2023} and references therein; 
(18) \citet{Hily-Blant2017} and references therein.

\end{sidewaystable*}

\end{appendix}

\end{document}